\def\vsk{\vskip.1in \noindent}

\hsize3.5in
\font\title=cmssdc10 scaled 1200
\magnification=\magstep1
\baselineskip 22 true pt \parskip=0pt plus3pt
\hsize 5.5 true in \hoffset .125 true in
\vsize 8.5 true in \voffset .1 true in

 \def\Schrod{Schr\"{o}dinger }
\def\pmb#1{\setbox0=\hbox{#1}
 \kern-.025em\copy0\kern-\wd0
\kern.05em\copy0\kern-\wd0
 \kern-.025em\raise.0433em\box0 }

\def\ket#1{|#1\rangle}

\def\ref#1{$^{\bf #1}$}
\def\bra#1{\langle #1|}

\def\kvac{\ket{0}}
\pageno=1
\def\bvac{\bra{0}}

\def\eq#1{\eqno{(#1)}}
\font\germ=eufm10
\centerline{{\title Disappearance of the Measurement Paradox}}
\centerline{{\title in a Metaplectic Extension of Quantum 
Dynamics}}
\vskip.2in
\centerline{Daniel I. Fivel}
\centerline{Department of Physics,
University of Maryland}
\centerline{College Park, Md.}
\vsk
\centerline{November 20, 2003}
\vskip.2in
\noindent
\centerline{Abstract}
\vskip.2in
\noindent
It is shown that \Schrod dynamics can be embedded in a larger
dynamical theory which extends its symmetry group  from
the unitary group to the full metaplectic group, i.e.\ the group of
linear canonical transformations.  Among the newly admitted
non-unitary processes are analogues of the classical measurement
process which makes it possible to treat the wave-function as an
objective property of the quantum mechanical system on the same
footing as the phase-space coordinates of a classical system. The
notion of ``observables" that in general have values only when
measured can then be dispensed with, and the measurement paradox
disappears.
\vskip.25in
\Schrod dynamics is restricted to unitary transformations.
  The
measurement paradox arises from the inadequacy of such
transformations to account for   measurement processes. Classical
dynamics, on the other hand, has no measurement paradox. The task of
classical measurement, which is to assign phase space coordinates to
an object system, can be performed through classically describable
interactions between object systems and  measuring devices. The
simplest such interactions (see below) are one-parameter subgroups
of the
 so-called metaplectic\ref{1} group ${\cal M}$ which is
 the group of linear canonical transformations, i.e.\
linear transformations of phase space that leave the Poisson bracket
invariant. However, they lie outside of its unitary subgroup ${\cal
M}_o$ and hence have no analogue in \Schrod dynamics. Thus while one
can construct a closed universe governed by classical mechanics, it is
not possible to construct a closed universe governed by the \Schrod
equation.
\vsk
In this paper it will be shown that \Schrod dynamics can be embedded
 in a larger system which extends the available
transformations from
${\cal M}_o$ to
${\cal M}$. A quantum mechanical analogue of the classical
measurement process then appears, and the
measurement paradox disappears. The resulting theory reproduces the
standard predictions within a closed universe.
\vsk
Let us begin by formulating \Schrod dynamics in a way that closely
resembles classical dynamics. Let $\Psi$ be a separable Hilbert space
of wave-functions, the components of which in some basis may be
written
$\psi_j = (p_j + iq_j)/\sqrt{2}$. The transformations of \Schrod
dynamics are one parameter subgroups of the unitary group which we
can write in the form
$$\psi \to \psi_t = e^{-itH}\psi,\eq{1a}$$
where $H$ is a hermitian matrix. Now compare this kind of
transformation of $\psi$ with what we would obtain if $\Psi$ were a
phase-space instead of a Hilbert space. We can pretend that the
$q_j$'s and
$p_j$'s are coordinates and momenta and subject $\psi$ to 
one-parameter sub-groups of the group of canonical transformations.
These are generated by
$C_{\infty}$ functions
$\Gamma$ of the
$p_j$'s and $q_j$'s or equivalently of the $\psi_j$'s and
$\psi_j^*$'s by the rule
$$\psi \to \psi_t = e^{-it Ad \Gamma}\psi,\eq{1b}$$
where $Ad$ acts by the Poisson bracket:
$$Ad{\cal A}\cdot B  = 
i\sum_k\left({\partial A\over\partial
q_k} 
{\partial B\over \partial p_k} - {\partial A\over\partial
p_k}{\partial B\over\partial q_k}\right) =
  \sum_k\left({\partial A\over\partial
\psi_k^*} 
{\partial B\over \partial \psi_k} - {\partial A\over\partial
\psi_k}{\partial B\over\partial \psi_k^*}\right).\eq{1c}$$
(For notational convenience this definition of the Poisson bracket
differs from the usual one by a factor of $i$.)
\vsk
The transformations of the metaplectic group ${\cal M}$ are
generated by {\it quadratic forms} $\Gamma(\psi^*,\psi)$ in the
components of
$\psi$ and $\psi^*$. They transform components of $\psi$ into linear
combinations of components of both $\psi$ and $\psi^*$ in general.
The subgroup ${\cal M}_o$ of ${\cal M}$ has generators of the form
$$ \Gamma = {\cal H}(\psi^*,\psi) =
\sum_{jk}\psi_j^*H_{jk}\psi_k,\eq{2}$$ where $H$ is a hermitian
matrix. These transformations transform components of $\psi$ linearly
without mixing in components of
$\psi^*$.
One then verifies that
$$e^{-it Ad {\cal H}}\psi_j = (e^{-itH}\cdot\psi)_j.\eq{3}.$$
Comparing this with (1a) we see that {\it all of
\Schrod dynamics can be described by one-parameter sub-groups of
${\cal M}_o$ acting on $\psi$ when we treat it as a point in a
phase-space rather than a vector in a Hilbert space.} This way of
representing \Schrod dynamics is not restricted to wave-functions
with discrete indices. For example a spatial wave-function $\psi(x)$
can be treated like a classical field with sums over indices 
replaced by integrals and the Poisson bracket defined with
variational derivatives.
\vsk
 The one-parameter subgroups of ${\cal M}$ that are
outside of
${\cal M}_o$ have no counterpart in \Schrod dynamics.  
They have generators  of the form 
$$\Gamma = {\cal W} = \sum_{jk}\psi_j^*W_{jk}\psi_k^* +
\hbox{complex  conjugate}.\eq{4}$$
These transformations transform components of $\psi$ and $\psi^*$
among one another. They are therefore linear only on real linear
combinations. 
We shall first show that processes
of this kind describe what happens in the simplest kind of
classical mesurement. This will motivate us to extend quantum dynamics
so that
$\psi$ can transform under the analogue of the full metaplectic group
thereby providing a way to incorporate quantum measurement into
the theory.
\vsk
In the simplest classical mesurement the object system and
measuring device will each have a single complex degree of freedom
with phase-space coordinates
$\lambda = (p + iq)/\sqrt{2}$ and $\mu = (P + iQ)/\sqrt{2}$
respectively. The task of the measurement
process is to {\it resolve} distinct object states and thereby
assign values of $\lambda$ to them. We assume that we know how to do
this for states of the measuring device provided that $|\mu|$ is
sufficiently large (macroscopic). We therefore seek an interaction
between object and device which, when strong enough, makes orbits of
$\mu$ for distinct initial $\lambda$ {\it diverge}
to any desired extent. Moreover it must do so in a time that can be
made arbitrarily small so that other dynamical processes that might
be present can be neglected.
\vsk  
Let us see how this can be accomplished. Consider the orbit of $\psi
= (\lambda,\mu)$ under the
 one-parameter metaplectic subgroup generated by
$${\cal W}  = \eta (\lambda\mu)^* + \eta^*(\lambda \mu),\eq{5}$$
where $\eta$ is a complex parameter. From (1b) we obtain:
$$\psi \to \psi_t =  e^{-it Ad {\cal W}}\psi =
(\lambda_t,\mu_t),\eq{6}$$
$$\lambda_t = \lambda\cosh(|\eta|t) +
\mu^*e^{i\arg\eta}\sinh(|\eta|t),$$ $$  \mu_t =
\mu\cosh(|\eta|t) +
\lambda^*e^{i\arg\eta}\sinh(|\eta|t).\eq{7}$$ 
To use this process to perform the task of assigning values to the
object coordinates, choose the initial device coordinate (the ``ready"
state) to be
$\mu = 0$ so that
$$\lambda_t = \lambda\cosh(|\eta|t),\;\;\mu_t =
\lambda^*e^{i\arg\eta}\sinh(|\eta|t),\eq{8a}$$
$$|\psi_t|^2 = |\lambda_t^2
+|\mu_t|^2| = |\lambda|^2\cosh(2|\eta|t).\eq{8b}$$
 Comparing the
$\mu$ orbits for two initial choices $\lambda^{(1)}$ and
$\lambda^{(2)}$ of
$\lambda$ we have
$$|\mu^{(1)}_t - \mu^{(2)}_t| =  |\lambda^{(1)} -
\lambda^{(2)}|\sinh(|\eta|t).\eq{8c}$$
 Thus no matter how close the initial values
$\lambda^{(1)}$ and $\lambda^{(2)}$ might be, we can choose $|\eta|$
sufficiently large that the distance between the device coordinates
at time t becomes as large as we please as quickly as we please. Thus
$\mu_t$ becomes a {\it macroscopic  pointer} from which (8a) gives
the initial value of $\lambda$ if $\eta$ is known. Since the process
is a group, there is an inverse by which the initial $\lambda$ can be
restored after the determination of $\mu_t$. Moreover, by choosing a
sufficiently large $|\eta|$, this can be done so rapidly that any
other dynamical processes that may be acting can be ignored. Thus
the availability of non-unitary metaplectic processes makes the
phase space coordinates a determinate property of a classical
system.
\vsk
Let us contrast this with quantum mechanics. In the
Dirac formulation we have the projective map from $\psi
\in
\Psi$ (excluding $\psi$ with $|\psi| = 0$) to unit state vectors
$$\psi
\to
\ket{\widehat{\psi}} \equiv \psi/|\psi|.\eq{9}$$
 There is a
non-vanishing probability that a system with state vector
$\ket{\widehat{\psi}}$  will pass a filter for a system with a
different state vector
$\ket{\widehat{\phi}}$ unless the two vectors are orthogonal.
  Thus only orthogonal states can be
perfectly resolved. To deal with this, the orthodox (Dirac-von
Neumann) theory introduces the notion of ``observables"
defined by hermitian operators. An observable does not have a
definite value until it is measured unless the system is in an
eigenstate. The  problems that arise from this interpretation, e.g.\
the unexplained collapse mechanism and the existence of grotesque
macroscopic superpositions (``\Schrod cats"), are
familiar\ref{2,3}. What is needed is, as J.S. Bell\ref{4} put it, a
theory of ``be-ables" rather than ``observables" in which objective
properties are assigned to systems.  
\vsk
 To treat $\psi$ as an assignable property of a classical system we
made use of the
 metaplectic  transformation that caused phase space distances
between distinct $\psi$'s to become arbitrarily large. Such
transformations have no quantum counterpart if we use the Dirac map
(9) because no change in the state vector $\ket{\psi}$ occcurs when
$\psi$ is multiplied by a scale factor $\tau > 0$. 
 The justification for (9) is that since
$|R\psi| = |\psi|$ for unitary transformations, 
$$\ket{R\psi} = R\ket{\psi}\eq{10}$$
which insures that \Schrod dynamics, which acts on
wave-functions and implements the superposition principle, transfers
properly to state vectors. Dirac \ref{5} defends the inability of (9)
to represent processes that scale the wave-function by asserting
that the norm
$|\psi|$ has no physical meaning. Our sought-after extension of
dynamics to include non-unitary processes will give $|\psi|$ physical
meaning and thereby justify modifying (9). Thus our first task is
to show that (9) can be replaced by a map that preserves (10) for
unitary transformations while permitting the representation of
processes in which the norm changes. 
\vsk 
For each $\psi$ in the Hilbert space $\Psi$ let there correspond a
unitary operator $U(\psi)$ on a Hilbert space {\germ h} such that
$U(0) = I$. Let $\kvac$ be a distinguished  unit vector of {\germ h}
which will be called the ``vacuum". The map
$$\psi \to \ket{\psi} \equiv U(\psi)\kvac \eq{11}$$
 will then define a unit vector associated with $\psi$. One may think
of
$\psi$ as the ``instructions" for creating the state from the vacuum.
\vsk
Let ${\cal F}$ be the set of unit vectors in {\germ h} that
correspond to 
 some $\psi \in \Psi$. 
Unlike Dirac kinematics, not every unit vector in
{\germ h} is necessarily a member of ${\cal F}$. The set ${\cal F}$
will be a manifold, but, since linear combinations of unitary
operators are {\it not} in general unitary,
{\it it will not be a linear space.} Hence there must be a
constraint on the form of
$U(\psi)$ to insure that the superposition principle, which holds in
the Hilbert space
$\Psi$, is properly implemented in  ${\cal F}$. Intuitively the
superposition principle says that we should be able to make the
state
$\ket{\psi^{(1)} +
\psi^{(2)}}$ from the operations 
used to make $\ket{\psi^{(1)}}$ and $\ket{\psi^{(2)}}$. This
suggests that $U(\psi)$ be required to satisfy
$$U(\psi^{(1)} + \psi^{(2)}) =
e^{i\theta_{12}}U(\psi^{(1)})U(\psi^{(2)})),\eq{12}$$ in which
the phase
$\theta_{12}$ may depend on the $\psi$'s and their order.
\vsk
Equation (12) is the defining relation for the Weyl-Heisenberg
group\ref{6}. We shall take it to be the fundamental condition
defining the symmetry of the theory, i.e.\ the group of allowed
transformations. A transformation
$\psi\to g\psi$, where
$g$ is not necessarily linear, will be allowed if (12) remains valid
(with the same phase) if $\psi$ is replaced by $g\psi$.
It suffices for this  that there exist a unitary
operator $V_g$ on {\germ h} such that
$$U(g\psi) = V_gU(\psi)V_g^\dagger.\eq{13}$$
Below we will explicitly construct $V_g$'s for one-parameter
subgroups of the metaplectic group, thereby
showing that (12) is consistent with an extension of dynamics to that
group.  
\vsk
The Stone-von Neumann theorem tells us that all of the
representations of (12) are obtained up to unitary equivalence (13)
by  tensoring  Fock representations together. Fock representations
are obtained by choosing a basis and exponentiating the Heisenberg
algebra, i.e.\
$$U(\psi) = e^{\psi\cdot a^\dagger - \psi^*\cdot a}\eq{14a}$$
in which 
$$\psi\cdot a^\dagger = \psi_1 a_1^\dagger + \psi_2 a_2^\dagger +
\cdots,\eq{14b}$$
where the $a$'s and their adjoints are operators on {\germ h} such
that:
$$ [a_i,a_j] = 0,\;\;\; [a_i,a_j^\dagger] =
\delta_{ij}I.\eq{15a},$$
 One then verifies that (12)
holds with
$$\theta_{12} = Im({\psi^{(1)}}^*\cdot \psi^{(2)}).\eq{15ba}$$
To define the state $\ket{\psi}$ we must specify the vacuum state
$\kvac$ which we take to be the tensor product of the states
annihilated by the $a_j$'s so that
$$a_j\kvac = 0 \;\; \forall j.\eq{16a}$$
It then follows that
$$\bvac\psi\rangle = e^{-|\psi|^2/2},\eq{16b}$$
whence from (12)
$$\bra{\psi^{(1)}}\psi^{(2)}\rangle = 
e^{i\theta_{12}}e^{-|\psi^{(1)}- \psi^{(2)}|^2/2}. \eq{16c}$$
Equation (16c) relates the  geometry of ${\cal F}$  to the geometry
of 
$\Psi$. It shows that {\it the effect on ${\cal F}$  
of successive dilation of the phase space $\Psi$ is to make
all of its vectors approach mutual orthogonality.} Observe that since
not all unit vectors of {\germ h} are in ${\cal F}$ there is 
``room" for the vectors of
${\cal F}$ to approach mutual orthogonality. 
\vsk
The appearance of bose operators in (14a) does not mean
  that this formalism applies only to systems with bose statistics.
For fermionic systems we need only restrict the Hilbert space $\Psi$
be a space of anti-symmetric wave-functions. Also, as remarked
earlier, one is not restricted to discrete indices. For example to
construct
$U(\psi)$ for states described by spatial wave-functions $\psi(x)$
we understand $\psi^*\cdot a$ to mean an integral over $x$ with
operators
$a(x)$ satisfying $[a(x),a(y)^\dagger] = \delta(x -
y)I$.
\vsk
We can  see quite simply why the metaplectic group preserves (12):
Since in general its elements transform components of $\psi$ linearly
into components of both $\psi$ and $\psi^*$, they act linearly on {\it
real} linear combinations such as the one that appears in the
argument of
$U$ on the left side of (12). Moreover one readily verifies that
to leave the Poisson bracket invariant the coefficients
$\alpha,\beta$ in
$\psi
\to
\alpha\psi + \beta\psi^*$ must also leave the imaginary part of the
scalar product 
$\theta_{12}$ (16a) invariant. The unitary subgroup ${\cal
M}_o$ leaves the scalar product itself invariant.
\vsk
 We can explicitly construct the
transformations $V_t$ that implement one-parameter subgroups $M_t$
of ${\cal M}$ as follows: Let $M_t$ act on $\psi$ by (1b) with
the quadratic generator $\Gamma(\psi^*,\psi)$.  The algebraic
relationship between commutator brackets and Poisson brankets is such
that we have the identity:
$$U(M_t\psi) = V_tU(\psi)V_t^\dagger,\eq{17a}$$
$$V_t = e^{-it:\Gamma(a,a^\dagger):}.\eq{17b}$$
Here $\Gamma(a,a^\dagger)$ is obtained by substituting $a$ for
$\psi^*$ and $a^\dagger$ for $\psi$ in the quadratic form
$\Gamma(\psi,\psi^*)$. The colons indicate normal ordering (putting
$a^\dagger$'s to the left of $a$'s). What makes (17) work is that
the transformation of (14a) by $V$'s of the form (17b)  transforms
the bose operators in the exponent  by 
$$a \to V a V^\dagger =  Aa + Ba^\dagger,\eq{18}$$
in which the matrices $A,B$ are such that the Heisenberg algebra
structure is preserved\ref{1}. 
\vsk
 We shall be interested in the orbit of the
quantum state corresponding to the orbit of $\psi$ under $M_t$. We
have
$$\ket{\psi_t} = U_t(\psi)\kvac,\;\; U_t(\psi) \equiv
U(M_t\psi).\eq{19}$$
 From (17) we have the
generalized \Schrod equation
$$\partial_tU_t(\psi) = -i Ad \Gamma\,
 U_t(\psi),\eq{20}$$
where $Ad$ acts by the commutator, i.e.\
$$Ad A\cdot B \equiv [A,B].\eq{21}$$
It reduces to the usual \Schrod equation when $\Gamma$ is a
generator of the unitary subgroup ${\cal M}_o$. To see this observe
that corresponding to the generator (2) we will have, according to
(17b), an operator
$\Gamma$ of the form
$${\cal H} = \sum_{jk} a_j^\dagger H_{jk}a_k,\eq{22}$$
which annihilates $\kvac$, so that $e^{-itH}$ leaves $\kvac$
invariant. Hence applying (17a) to $\kvac$ we obtain
$$\ket{e^{-iHt}\psi} = e^{-itH}\ket{\psi}.\eq{23}$$
Thus  the property (10) of the Dirac map needed for the
implementation of \Schrod dynamics is preserved.
 Indeeed we obtain the usual \Schrod equation 
$$\partial_t\ket{\psi_t} = - i {\cal H}\ket{\psi_t}\eq{24}$$
when transformations are restricted to the unitary subgroup ${\cal
M}_o$.
\vsk
We are now able to construct the quantum mechanical analogue
of the classical measurement process described in equations 6-8.
The generalization of (5) for a multi-dimensional phase-space is
$${\cal W} = \sum_j {\eta_j(\lambda_j\mu_j)^* + \hbox{complex
conjugate}}.\eq{25}$$
Assuming again that the ready state of the device corresponds
to $\mu = 0$ the orbit will be $\ket{\psi_t}$ with $\psi_t =
(\lambda_t,\mu_t)$ where 
$$\lambda_{jt} = \lambda_j\cosh(|\eta_j|t),\;\;
\mu_{jt} = \lambda_j^*e^{i\arg \eta_j}\sinh(|\eta_j|t).\eq{26}$$
Suppose that we let $|\eta_j|$ have the same value $|\eta|$
for all $j$. 
If $\psi^1_t$ and $\psi^2_t$ are the orbits corresponding to 
different initial values $\lambda^1$ and $\lambda^2$ of the object
system we have from (16c)  
$$|\bra{\psi^1_t}\psi^2_t\rangle|^2 = e^{-|\lambda^1  -
\lambda^2|^2\cosh(2|\eta|t)}\eq{27}$$
 which tends to zero in a super-rapid way since a
hyperbolic function appears in the exponent. 
 Given any lattice on the
object phase space, no matter how fine, and any $\epsilon > 0$, we
can choose a sufficiently large $|\eta|$ that all scalar products
(27) between states corresponding to distinct lattice points become
smaller than $\epsilon$ at a time $t$ which can be made arbitrarily
small. Thus the effect of the interaction is to make the set of state 
vectors associated with distinct $\lambda$'s on the lattice become
mutually orthogonal to any desired approximation. The set thus 
becomes classical in the sense that all propositions have only yes-no
answers except for fluctuations of negligible probability. Thus as
in the classical case we can in principle read the coordiates
$\mu_{jt}$ for all $j$ of the device and thereby deduce $\lambda_j$
from the relation
$$\mu_{jt} = e^{i\eta_j}\lambda_j^*\sinh(|\eta|t)\eq{28a}$$
provided that $|\eta|$ and $\arg \eta_j$ are known. If we do not
know $|\eta|$ or the phases of $\eta_j$ we can nonetheless determine
the ratios
$$|\lambda_j/\lambda_k|^2 = |\mu_{jt}/\mu_{kt}|^2.\eq{28b}$$
 Being a group, the process
generated by 
${\cal W}$  is reversible, so we can assign initial $\lambda$-values
to object states and return to those states in a short enough time
that the effect of other dynamical processes that might be present
can be neglected. 
\vsk
We have now established that the metaplectic extension of \Schrod
dynamics introduces non-unitary processes by which the wave-function
of an object system becomes a determinate property of the system on
the same footing as the phase-space coordinates of a classical
system. In the extended theory the norm $|\psi|$ has physical
meaning which we explore next.
\vsk
Let us first observe that there is a sense in which the Dirac map
(9) is the limiting form of (11) for $|\psi|
\to 0.$ To see this note that when
 (9) is used there can be no state corresponding to $\psi \equiv
0$ whereas with (11) this corresponds to the vacuum $\kvac$.  We
therefore examine  the normalized projection 
 $\ket{\psi_\perp}$ of $\ket{\psi} $ in the
direction orthogonal to the vacuum. From (16b) one obtains
$$\ket{\psi_\perp} = (\ket{\psi}
- e^{-|\psi|^2/2}\kvac)/(1 - e^{-|\psi|^2})^{1/2}.
\eq{29}$$
From (11) and (14a) we have
$$ \ket{\widehat{\psi}} \equiv    \lim_{|\psi| \to 0}\ket{\psi_\perp}
=  (\widehat{\psi}\cdot a^\dagger)\kvac ,\;
\;\;\widehat{\psi} \equiv
\psi/|\psi|.\eq{30}$$
Thus 
$$\bra{\widehat{\phi}}\widehat{\psi}\rangle =
\widehat{\phi}^*\cdot\widehat{\psi},\eq{31}$$
which identifies the states $\ket{\widehat{\psi}}$ as the Dirac
states defined by (9). Hence
$$\ket{\psi_\perp} \to
\cases{\ket{\widehat{\psi}},& $|\psi| \to 0$
\cr \ket{\psi},&$|\psi|\to \infty$},\eq{32}$$
so that $\ket{\psi_\perp}$ {\it interpolates between 
states with small $|\psi|$ that act
quantum mechanically and  states with large $|\psi|$ that act
classically.} 
Thus the norm $|\psi|$ is a
measure of the ``classicality" of the system.
\vsk
It follows from (12) that for any integer $n$ 
$$U(\psi) = (U(\psi/n))^n\eq{33}$$
so that the states with large norm can be created by repeated
application of the creation operator $U$ for a state with an
arbitrarily small norm. This suggests that  $U(\psi)$ can be
interpreted as the creator of a ``beam" of copies of the quantum
state defined by
$\widehat{\psi}$, with an intensity given by some monotonically
increasing function
$I(|\psi|)$ of the norm. We can deduce this function as
follows: If $\psi^a,\psi^b$ belong to orthogonal subspaces of
$\Psi$ so that  $U(\psi^a)$ and $U(\psi^b)$  commute, there
will be no interference when the beams they create are combined.
Hence the intensities simply add. Thus from (12)
$$I(|\psi^a + \psi^b|) = I(|\psi^a|) + I(|\psi^b|),\eq{34}$$
whence except for an arbitrary choice of scale we must set
$$I(|\psi|) = |\psi|^2.\eq{35}$$
\vsk
In the standard description of quantum measurements each complete
set of commuting observables defines a basis in $\Psi$ considered as
a Hilbert space, namely the basis in which they are all diagonal.
This determines a decomposition 
$$\psi = \sum_j\psi^j\eq{36}$$
in which $\psi^j$ is the projection of $\psi$ on the one-dimensional
subspace determined by the $j$'th basis vector. It follows from
(12) that the operators $U(\psi^j)$ mutually commute and that
$$U(\psi) = \prod_j U(\psi^j).\eq{37}$$
Thus each complete set of commuting observables defines a
factorization of $U(\psi)$ in which the factors represent
non-interacting beams. The intensity of the beam created by
$U(\psi)$ will be the sum of the intensities of the constituent
beams.
\vsk
Corresponding to the factorization (36) there will be a
factorization of the vacuum state into a tensor product of
states $\ket{0,j}$ annihilated by $a_j$. Defining
$$N_j =  a_j^\dagger a_j,\;\; N = \sum_j N_j,\eq{38}$$ we see
that the $N_j$'s 
form a complete set of commuting observables. 
From (14,15)
$$a_j\ket{\psi} = \psi_j\ket{\psi},\eq{39}$$
whence we obtain the expectation value
$$ \overline{N_k} \equiv \bra{\psi_\perp}N_k\ket{\psi_\perp} =
|\psi_k|^2(1 - e^{-|\psi|^2})^{-1} \to 
\cases{|\widehat{\psi}_k|^2,& $|\psi| \to 0$,\cr
|\psi_k|^2, & $|\psi|\to \infty$}.\eq{40}$$
Thus in the quantum limit $\overline{N_k}$ can be interpreted as the
probability of a copy being in the $k$'th subbeam whereas
in the classical limit it coincides with the intensity of
that subbeam. 
\vsk
Let \ $\Delta N_k$ be the 
dispersion of $N_k$ in the  state $\ket{\psi_\perp}$.
Then one verifies that
$$\Delta N_k/\overline{N_k} \to 1/|\psi_k| \;\hbox{ for } \;|\psi|
\to
\infty.\eq{41}$$
This goes to zero for any $k$ for which
$\widehat{\psi_k} \neq 0$. Thus there is a sharp value of  $N_k$,
in the $k$'th subbeam, namely the intensity
$|\psi_k|^2$ of the subbeam. Ratios of these intensities then give
the ratios
$|\widehat{\psi_k}/\widehat{\psi_j}|$. As we saw in (28b) it is these
ratios that we can obtain in the quantum analogue of the
classical measurement process when we cannot control the
amplification parameter $\eta$. Thus if we knew how to implement the
amplification process generated by ${\cal W}$ but could not control
$\eta$, the information we would obtain about the state would be
identical to the information obtainable by comparing subbeam
intensities when the intensities are large. This is precisely what 
we predict from quantum measurements as they are described in the
orthodox formulation.
\vsk
The operator $N$
commutes with all generators of the form (22) and therefore defines
an observable that is constant for all \Schrod processes.
The expectation values of $N$ and $N^2$ in $\ket{\psi_\perp}$ are
given by
$$\overline{N} = |\psi|^2(1 - e^{-|\psi|^2})^{-1},\;\;
\overline{N^2} = (|\psi|^4 + |\psi|^2)(1 -
e^{-|\psi|^2})^{-1}.\eq{42}$$ 
Note that
$$\overline{N} \to 
\cases{1,& $|\psi| \to 0$,\cr
|\psi|^2, & $|\psi|\to \infty$}.\eq{43}$$
Thus if we interpret $N$ as a ``counter" for the number of copies in
the beam, the quantum limit has one copy while the classical limit
has a large number indicated by $|\psi|^2$. If
$\Delta N$ is the dispersion we then find from (42) that
$\Delta N/\overline{N}$ tends to zero both for  $|\psi| \to 0$ and
$|\psi| \to \infty$.
 It has a maximum of $\approx 0.55$ at $|\psi| \approx 1.8.$
Thus this ratio provides a ``marker" for the transition from the low
intensity quantum mechanical regime to the high intensity classical
regime. 
\vsk
We have now shown that \Schrod dynamics can be embedded in a larger
framework that enjoys the full metaplectic group as its symmetry
group and that the extension incorporates a quantum analogue of 
the classical measurement process. The wave-function $\psi$ 
becomes a determinate property of systems such that those with
small $|\psi|$ behave quantum mechanically and those with large
$|\psi|$ behave classically. In the latter case $|\psi|^2 $
is the intensity of a beam consisting of copies of the state
described by the unit vector $\widehat{\psi}$.
\vsk
 The processes of the extended
theory obey a generalized \Schrod equation which reduces to the usual
\Schrod equation for unitary processes, so that the predictions of
the standard theory are unaltered. It is no longer necessary to
formulate the theory in terms of observables which only have values
when they are measured. Hence the measurement paradox disappears, and
it is possible to construct a closed universe governed by the
generalized \Schrod equation.
 
\vsk
\centerline{References}
\vsk
1. G.B. Folland (1989), {\it Harmonic Analysis in Phase Space},
(Princeton: Princeton University Press).
\vsk
2. J. Bub (1997), {\it Interpreting
the Quantum World}, (Cambridge: Cambridge University Press).
\vsk
3. A.J. Leggett (1987), {\it Quantum Implications}, B.J. Hiley and F.
David Peat, ed. Chap. 5, (London and N.Y: Rutledge).
\vsk
4. J.S. Bell (1987), {\it Speakable and Unspeakable in Quantum
Mechanics} (Cambridge: Cambridge University Press).
\vsk 
5. P.A.M. Dirac (1958), {\it The Principles of Quantum Mechanics}, 
p.17,$\,$ (Oxford: Clarendon Press).
\vsk
6. A. Perelomov (1986),
 {\it Generalized Coherent States and Their Applications}, Sec.
1.1, (Berlin: Springer-Verlag).
\end